# Zeolite-based photocatalysts immobilized on aluminum support by plasma electrolytic oxidation

Kristina Mojsilović[a], Nikola Božović[a,b], Srna Stojanović[c], Ljiljana Damjanović-Vasilić[c], Maria Serdechnova[d], Carsten Blawert[d], Mikhail L. Zheludkevich[d,e], Stevan Stojadinović[a], Rastko Vasilić[a,*]

[a] University of Belgrade, Faculty of Physics, Studentski trg 12-16, 11000 Belgrade, Serbia

[b] Directorate of Measures and Precious Metals, Mike Alasa 14, 11000 Belgrade, Serbia

[c] University of Belgrade, Faculty of Physical Chemistry, Studentski trg 12-16, 11000 Belgrade, Serbia

[d] Institute of Surface Science, Helmholtz-Zentrum Hereon, Max-Planck-Straße 1, 21502 Geesthacht, Germany

[e] Institute of Materials Science, Faculty of Engineering, Kiel University, Kaiserstraße 2, 24143 Kiel, Germany

*Corresponding author. Tel: + 381-11-7158161; Fax: + 381-11-3282619

E-mail address: rastko.vasilic@ff.bg.ac.rs (Rastko Vasilić)

**Abstract**

The preparation and properties of zeolite-containing oxide coatings obtained by plasma electrolytic oxidation are investigated and discussed. Pure and Ce-exchanged natural (clinoptilolite) and synthetic (13X) zeolites are immobilized on aluminum support from silicate-based electrolyte. Obtained coatings are characterized with respect to their morphology, phase and chemical composition, photocatalytic activity and anti-corrosion properties. It is observed that all mentioned properties of obtained coatings are dependent on processing time and type of immobilized zeolite. Coatings with Ce-exchanged zeolite show higher photocatalytic activity and more effective corrosion protection than those with pure zeolite. The highest photocatalytic activity is observed for coatings processed in pulsed a DC regime for 30 minutes containing Ce-exchanged 13X zeolite, followed by those containing Ce-exchanged clinoptilolite. Pronounced anti-corrosion properties feature almost all samples containing Ce-exchanged 13X zeolite.

**1. Introduction**

The removal of contaminants from wastewater and helping to save water resources are major topical questions driving numerous researchers around the world. Photocatalytic degradation under visible light irradiation is considered to be an advanced technology in many fields of waste treatment, and for that purpose many semiconducting photocatalysts have been developed [1-4].

Photocatalytic process is usually carried out in a batch slurry photoreactor operating with nanoparticle suspensions [5-7] which result in numerous practical and economical disadvantages [8]. In contrast, utilization of immobilized oxide coatings reduces the net amount of catalyst surface available to photocatalysis, consequently decreasing the efficiency of the process [9]. To enhance the efficiency of the process, significant efforts have been directed towards the synthesis of composite photocatalysts, where plasma electrolytic oxidation (PEO) proved to be an appropriate preparation technique for production of stable oxide coatings [10].

PEO is a well-established industrial surface treatment method that can be used to convert the surface of a number of metals (Al, Mg, Ti, Zn and their alloys) to their oxides with possible incorporation of additional elements into obtained coatings through modification of electrolyte composition [11]. A number of articles were focused on the incorporation of particles into oxide coatings during the PEO treatment [12-14]. Generally, addition of particles to electrolyte used for PEO processing can act as a sealant making obtained oxide coatings denser, thus increasing their wear and corrosion resistance. It has also been shown that incorporation of particles influences the onset of dielectric breakdown, chemical and phase composition of obtained oxide coatings and enhances luminescent and photocatalytic properties [13,15].

Currently a lot of attention is focused on possible functional additives to PEO electrolytes and zeolites seem like a viable candidate for this purpose. Zeolites are microporous crystalline materials with a three-dimensional framework composed of pores and channels of regular dimensions. Conventional zeolitic frameworks are composed of $SiO_4$ and $AlO_4$ tetrahedra which are linked together by sharing oxygen atoms. Presence of Al leads to a negative charge on the framework compensated by presence of cations in the pore structure. The metal ions in the pore structure can be easily replaced by other cations through ion exchange process [16]. The ion exchange process is affected by structure type, Si/Al ratio, location of cation in the zeolitic structure, hydration degree and the preparation conditions (temperature, concentration of exchangeable cation, presence of the other exchangeable cations, type of anions).

Clinoptilolite is an abundant and inexpensive naturally occurring zeolite, characterized by framework of parallel channels formed by eight- and ten-membered tetrahedral rings (3.3×4.6 and 3.0×7.6 Å, respectively) intersecting with channel composed of eight-membered tetrahedral rings (2.6×4.7 Å) [17,18]. Synthetic zeolite X is constructed from sodalite cages (truncated cuboctahedron with tetrahedral Si or Al atoms located at the 24 corners) connected by double six-membered rings forming large pores known as supercages with a diameter of ~12.6 Å. The sodalite cages are arranged tetrahedrally in the structure of zeolite X resulting in three orthogonal small pores of ~7.5 Å diameter [19].

Regardless of the fact that zeolites are mainly used as adsorbents it has been shown that they could enhance the efficiency and selectivity of photocatalysts either by photoactivating the zeolite framework or by encapsulating nano-sized semiconductor oxides [20-23]. Among semiconducting oxides $CeO_2$ is an excellent candidate for the photocatalytic decomposition of organic pollutants even surpassing $TiO_2$ in certain applications [24-27]. At the same time, a number of studies focused on corrosion protection revealed that Ce-exchanged zeolites may be successfully used for this purpose [28-30].

The aim of this work is to investigate the coatings with immobilized pure and Ce-exchanged clinoptilolite and 13X zeolites on aluminum support using plasma electrolytic oxidation processing. Obtained coatings are characterized with respect to their surface morphology, phase and chemical composition. Photocatalytic properties of obtained coatings are probed by decomposing methyl orange under simulated sunlight radiation and diffuse reflectance spectroscopy. Coatings are also tested for their corrosion properties using electrochemical impedance spectroscopy.

## 2. Materials and methods

*2.1 Preparation of zeolites*

The starting zeolites used in this work were synthetic FAU-type zeolite Na-13X ($Na_{87}[Al_{87}Si_{105}O_{384}]$, Si/Al = 1.2) produced by Union Carbide and HEU-type natural zeolite clinoptilolite (($Na,K)_6Si_{30}Al_6O_{72}·nH_2O$, Si/Al = 5) originated from Zlatokop mine in Vranjska Banja, Serbia. Cerium containing forms of zeolites were obtained using conventional aqueous ion exchange procedure in dilute solution [31]: 5 g of each zeolite were suspended in 1 L of 0.003 M $Ce(NO_3)_3·6H_2O$ (purity 99 %, supplied by Aldrich) solution, then stirred for 7 days at room temperature, followed by filtering, rinsing with deionized water and drying overnight at 80 ºC in air. The exchange process was repeated twice. The cerium content reached to 25.5 ± 1.3 wt% in the case of 13X zeolite and 2.0 ± 0.6 wt% in the case of clinoptilolite.

*2.2 PEO processing*

Rectangular samples of 1050 grade aluminum alloy were set as anode and used as the support for zeolite immobilization. Stainless steel sheet of approx. 20 $cm^2$ was used as cathode in all experiments. The anode material was sealed with insulation resin leaving an active surface area of approx. 2.5 $cm^2$ accessible to electrolyte. Aqueous solution of 4 g/L $Na_2SiO_3$ + 4 g/L KOH was used as supporting electrolyte for all experiments, with additions of 1 g/L of clinoptilolite, clinoptilolite exchanged with Ce, 13X, and 13X exchanged with Ce zeolites (Table 1).

**Table 1.** Sample designation and electrolyte used for PEO processing.

| Sample | Electrolyte |
|---|---|
| Cli | 4 g/L $Na_2SiO_3$ + 4 g/L KOH +1 g/L clinoptilolite |
| Cli+Ce | 4 g/L $Na_2SiO_3$ + 4 g/L KOH +1 g/L clinoptilolite+Ce |
| 13X | 4 g/L $Na_2SiO_3$ + 4 g/L KOH +1 g/L 13X |
| 13X+Ce | 4 g/L $Na_2SiO_3$ + 4 g/L KOH +1 g/L 13X+Ce |
| SE | 4 g/L $Na_2SiO_3$ + 4 g/L KOH |

Electrolyte was prepared using double distilled and deionized water and p.a. (pro analysis) grade chemical compounds. PEO process was carried out in a jacketed electrolytic cell maintaining the temperature of the electrolyte below 30 °C. During PEO processing all electrolytes with zeolite addition were agitated by a magnetic stirrer. A home-made pulsed DC power supply working in galvanostatic (current controlled) mode was used for this experiment. The power supply produced rectangular pulses of 2.5 A with a possible $t_{on}$ value ranging from 1 to 20 ms and a $t_{off}$ value from 1 to 2 s [32]. PEO processing was performed under a current density of 1 $A/cm^2$ during 10, 20, 30 and 40 min with a $t_{on}$ of 10 ms and $t_{off}$ of 1 s for all experiments (Figure 1). Voltage pulse sequences were recorded using Tektronix TDS 2022 digital storage oscilloscope and a high voltage probe which was connected directly to the power supply.

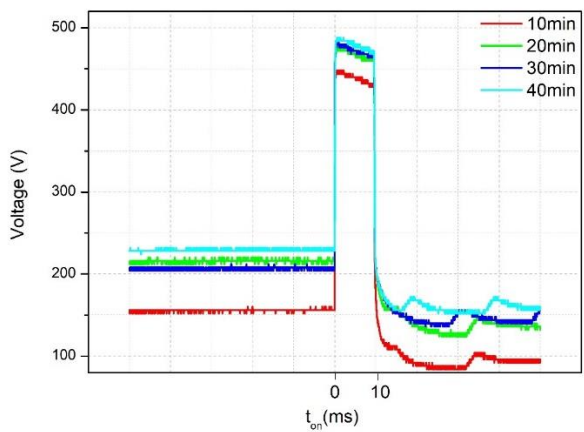

**Figure 1.** Representative waveform used in all experiments and its change with PEO processing time.

*2.3 Characterization*

A scanning electron microscope (SEM, Tescan Vega3 SB) in BSE UniVac mode was used to examine surface morphology of the PEO layers, while chemical composition of their surface was analyzed with energy dispersive spectrometer (EDS, eumeX) coupled with SEM. Porosity and surface roughness of obtained coatings were estimated using ImageJ software. For cross-sectional SEM analyses, samples were embedded in epoxy resin and polished with 220, 1000, and 4000 SiC abrasive papers, followed by polishing with 1 μm diamond paste. Cross-sectional micrographs were acquired using JEOL 840A with SE detector.

Rigaku Ultima IV diffractometer with Ni-filtered CuK$_α$ radiation source was used for crystal phase identification. Crystallographic data was collected in Bragg-Brentano mode, in $2\theta$ range from 20° to 70° with a scanning rate of 2 °/min.

Photoluminescence (PL) spectral measurements were taken on a Horiba Jobin Yvon Fluorolog FL3-22 spectrofluorometer at room temperature, with a 450 W xenon lamp as the excitation light source. Obtained spectra were corrected for the spectral response of the measuring system and spectral distribution of excitation light source.

Photocatalytic activity of obtained coatings on Al substrate was determined by degrading methyl orange (MO) at room temperature. Samples were immersed into 10 mL of 8 mg/L aqueous MO solution and placed on a perforated holder with a magnetic stirrer underneath. Prior to irradiation, the solution and the catalyst were magnetically stirred in the dark for corresponding irradiation time to check coatings' adsorption properties. For irradiation Osram Vitalux lamp (300 W) that simulates solar spectrum was placed 25 cm above the top surface of the solution (corresponding to illumination intensity of about 16000 lx). A fixed quantity of the MO solution was removed every hour to measure the absorption and then to determine concentration using UV–Vis spectrophotometer Agilent Carry 60. After each measurement probe solution was returned back to the photocatalytic reactor. Prior to photocatalysis measurements, MO solution was tested for photolysis in the absence of the photocatalyst in order to examine its stability. The lack of change in the MO concentration after 6 h of irradiation revealed that degradation was only due to the presence of photocatalyst.

UV–Vis diffuse reflectance spectra (DRS) of oxide coatings were recorded using UV–Vis spectrophotometer Agilent Carry 5000 equipped with an integrating sphere.

The degradation behavior of PEO treated specimens (and bare substrate for comparison) was investigated via electrochemical impedance spectroscopy (EIS) measurements using a Gill AC potentiostat (Gill AC, ACM Instruments, UK) in 3.5 wt.% NaCl solution (ca. 300 mL). Traditional three-electrode cell containing Ag/AgCl reference electrode, platinum counter electrode and specimen as a working electrode (exposed area of 0.5 cm$^2$) was used. All electrochemical tests were performed under normal atmosphere at room temperature (ca. 21 °C) using magnetic stirring (ca. 200 rpm). The EIS measurements were performed vs. OCP with an AC amplitude of 10 mV (RMS) over a frequency range of 30000 Hz–0.5 Hz (overall 70 data points were recorded). The impedance spectra were collected after immersion for 5 min, 1 h, 3 h, 6 h, 12 h, 24 h, 48 h, 72 h, 96 h, 120 h, 144 h and 168 h in order to study the degradation process as a function of immersion time.

## 3. Results and discussion

*3.1 Crystallinity, morphology, and chemical composition of obtained coatings*

Survey XRD patterns of oxide coatings on Al containing clinoptilolite obtained for various PEO processing times are presented in Fig. 2. Presented XRD patterns are representative for all electrolytes used in this study, i.e., features on them are related only to crystalline phases originating from the aluminum substrate (denoted as S) and supporting electrolyte. XRD maxima originating from pure and Ce-exchanged zeolites are not present in these diffractograms, most probably as a result of their low concentration and/or good dispersion in obtained PEO coatings. Prolonged PEO processing times result in a decrease of reflections originating from the substrate and an increase of the reflection denoted as γ+Sil (γ-$Al_2O_3$+sillimanite) in diffractograms.

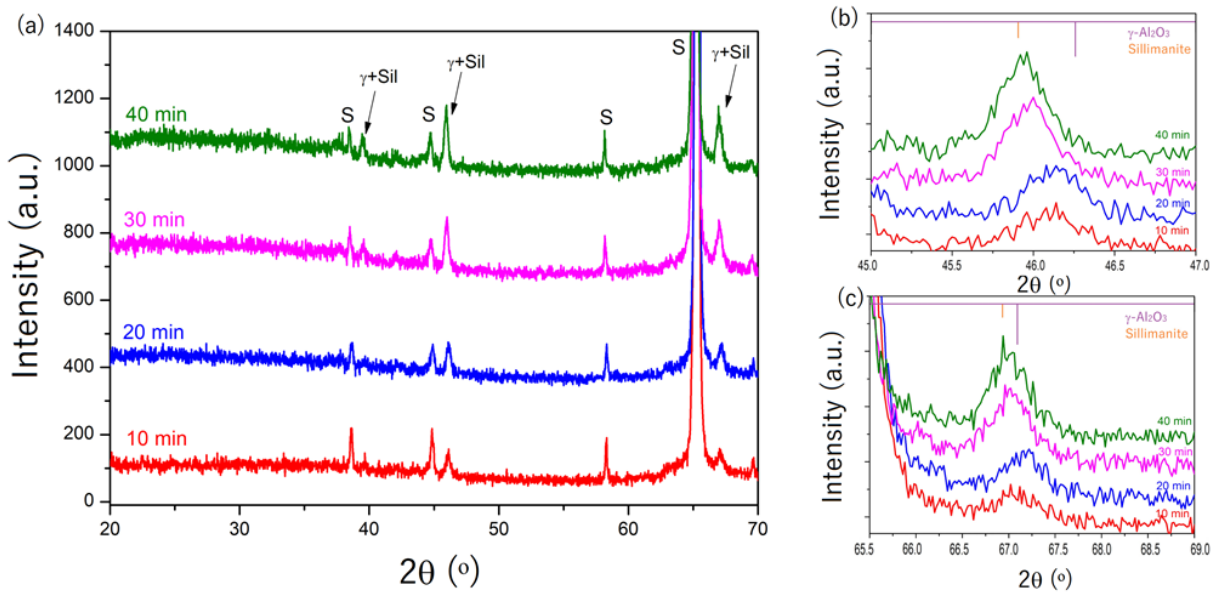

**Figure 2.** a) XRD patterns of oxide coatings on Al containing clinoptilolite obtained for various PEO processing times; b) and c) high resolution XRD patterns in the range of interest.

A closer inspection of diffraction maxima increasing with PEO processing time (Fig. 2b and Fig. 2c) show that at the beginning of PEO processing this peak can be attributed to crystalline γ-alumina (JCPDS 10-0425) phase, while at later processing times it shifts towards aluminosilicate sillimanite (JCPDS 22-0018) crystalline phase. Therefore, the observed relatively wide peak is a result of coexistence of the two crystalline phases. At the beginning of PEO processing with generous supply of $Al^{3+}$ ions from the substrate γ-alumina peak forms first while later on oxide coating thickens and with increased supply of silicate species from the electrolyte sillimanite diffraction maximum dominates.

In our recently published study [32] we were not able to observe diffraction peaks corresponding to α-alumina phase in obtained coatings using the same power supply. This may indicate that the magnitude of the applied current density is not sufficient to support this phase transformation. Although we used reasonably high current density (1 $A/cm^2$) with corresponding voltage of about 500 V (see Fig. 1), these conditions are applied for a very short time (i.e. 10 ms), which probably shuts down the micro-discharges quickly and prevents phase transformation during the PEO process. Despite locally high temperatures inherent to PEO, the absence of α-alumina was also observed in similar system under pulsed unipolar PEO processing conditions [33].

The surface morphology of coatings obtained for various PEO processing times in supporting electrolyte containing 1 g/L of Ce-exchanged 13X zeolite are presented in Fig. 3. SEM micrographs show the usual cracked pancake-like morphology decorated by dispersed beads attached to the coating surface, typical of the PEO coatings on Al alloys [34, 35]. With longer PEO processing times, the dielectric breakdown of alumina layer occurs at flaws formed in the oxide layer creating discharge channels from substrate to coating/electrolyte interface, resulting in larger diameter of micro-discharging channels [36, 37]. The current flowing through the breakdown sites causes melting and vaporizing of both substrate and already formed oxide coating. The molten material moves through the micro-discharge channels and quenches upon contacting the electrolyte, resulting in the formation of solidified molten regions (pancakes) around the discharge channels.

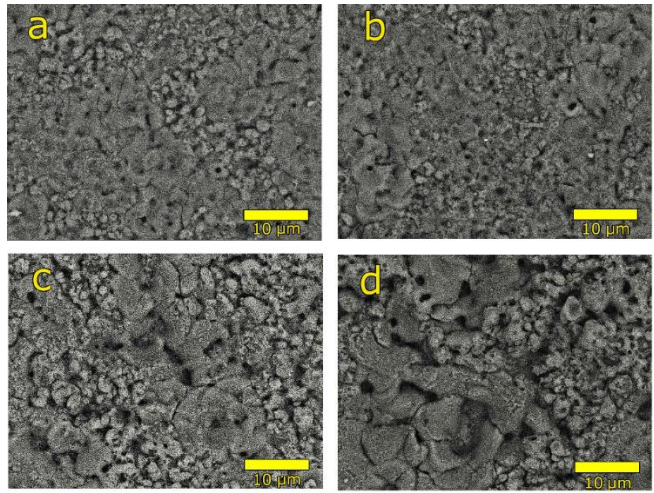

**Figure 3.** Top-view micrographs of coatings obtained in 4 g/L $Na_2SiO_3$ + 4 g/L KOH + 1g/L 13X+Ce after: a) 10 min; b) 20 min; c) 30 min; d) 40 min.

To quantify the observed change in the surface morphology of obtained coatings roughness and porosity data were extracted from corresponding micrographs (Table 2). Obviously, roughness increases up to 30 min of PEO processing time for all electrolytes used in this study, and then it levels off or slowly decreases. On the other hand, porosity of obtained coatings shows highest values for 10 min processing time and after that decreases. Data from Table 2 can be related to the influence of PEO processing time on coatings' morphology. Namely, longer PEO processed coatings are thicker and requiring higher energy for each dielectric breakdown. As a result, the current is localized at flaws in the oxide layer working its way through the oxide coating, but leaving behind a discharge channel of a larger diameter. As observed from visual inspection of SEM micrographs, the number of the micro-discharging channels decreases while the size of the pores observable on the coating's surface increases with PEO processing time.

**Table 2.** Estimated surface roughness and porosity data for various PEO processing parameters.

| Sample | PEO time [min] | $R_a$ [μm] | Porosity (%) |
|---|---|---|---|
| 13X | 10 | 2.23 | 19.02 |
|  | 20 | 2.24 | 16.42 |
|  | 30 | 2.79 | 14.33 |
|  | 40 | 2.81 | 16.04 |
| 13X+Ce | 10 | 2.36 | 16.39 |
|  | 20 | 3.07 | 13.47 |
|  | 30 | 3.43 | 11.05 |
|  | 40 | 3.46 | 12.14 |
| Cli | 10 | 1.55 | 21.88 |
|  | 20 | 2.21 | 19.69 |
|  | 30 | 2.45 | 16.34 |
|  | 40 | 2.49 | 17.12 |
| Cli+Ce | 10 | 1.84 | 20.03 |
|  | 20 | 2.49 | 17.36 |
|  | 30 | 3.29 | 13.42 |
|  | 40 | 3.30 | 16.25 |

Representative cross-sections of the oxide coatings formed after 20 min of PEO processing are shown in Fig. 4. Except for the coating obtained in supporting electrolyte alone (Fig. 4a), obtained coatings are continuous (thickness ~7 µm) and feature well developed porosity and roughness, as outlined in Table 2. Coating thickening and compaction is a result of addition of zeolite particles to the supporting electrolyte and their probable stimulated reactive incorporation into oxide coatings due to the low melting point [13]. Despite the fact that majority of published PEO related studies render porous and rough coatings non-desirable, this is not always true. Thick, dense, non-porous coatings of low surface roughness are desirable in the field of corrosion protection, while coatings with developed large surface area are of interest in the fields of catalysis and luminescence [38].

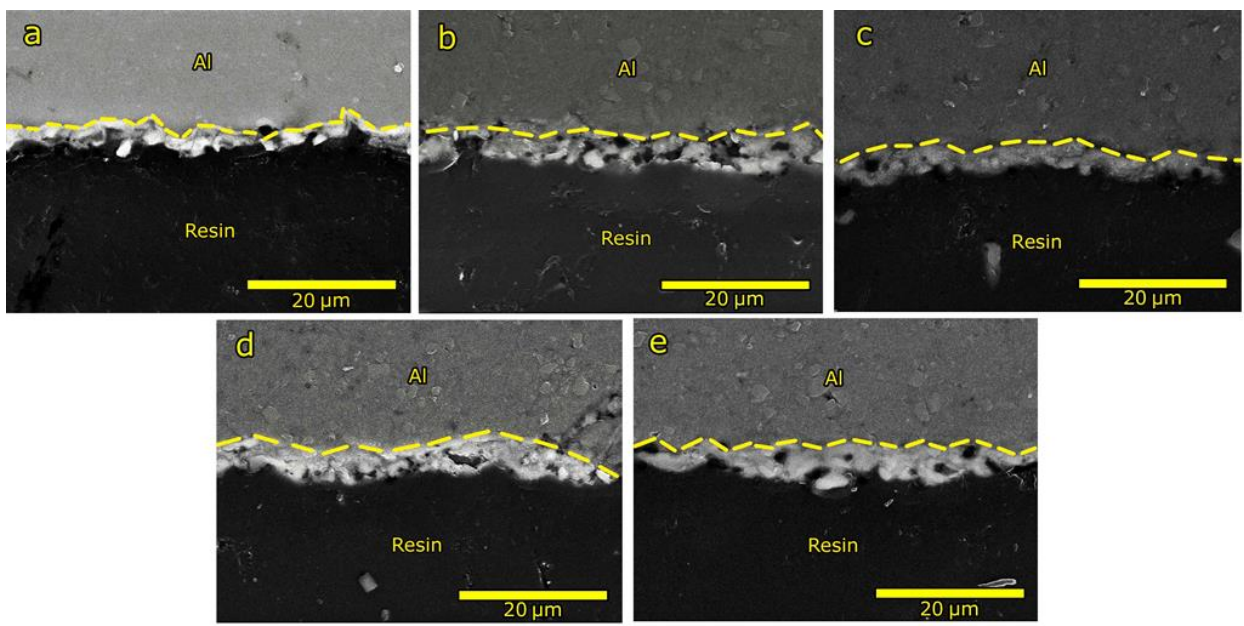

**Figure 4.** SEM micrographs of oxide coatings cross sections formed for 20 min: a) SE; b) Cli; c) Cli+Ce; d) 13X; e) 13X+Ce.

Top surface of the coatings was used for investigating the chemical composition of formed coatings. EDS data (as weight percentage (wt. %)) of the elements present in the coatings is shown in Table 3. The main elements of the coatings are Al, O, Si, Na, K, Fe, and Ce indicating the contribution of elements from both substrate and electrolyte. The presence of Fe stems from the fact that it is the most common impurity in zeolites, while observed C is related to utilization of carbon tape used for mounting the samples onto holder and from the vacuum system.

**Table 3.** EDS analysis of the surface coatings obtained after 40 min of PEO processing in various electrolytes.

| Element | Sample | | | | |
| --- | --- | --- | --- | --- | --- |
| | Cli | Cli+Ce | 13X | 13X+Ce | SE |
| | Weight percent (wt %) | | | | |
| Al | 49.11 | 50.24 | 50.37 | 45.50 | 37.07 |
| Si | 4.98 | 5.30 | 5.52 | 6.61 | 8.40 |
| O | 38.90 | 38.82 | 38.94 | 37.72 | 33.83 |
| K | 0.50 | 0.99 | 0.90 | 1.21 | 3.60 |
| Na | 0.77 | 0.97 | 0.97 | 0.85 | 1.03 |
| Fe | 0.14 | 0.15 | - | - | - |
| C | 5.55 | 3.45 | 3.27 | 7.91 | 16.07 |
| Ce | 0.05 | 0.08 | 0.03 | 0.20 | - |

However, EDS elemental composition results should be taken with precaution, because the accuracy of elemental composition determination varies for elements under investigation from 0.1 % for Ce to 8 % for O (given as standard deviation by the instrument). In the case of oxide coatings obtained by PEO it is suggested that accuracy of ± 5 % should be used [39]. Having this in mind it is evident from Table 3 that the content of all elements is almost the same for all coatings. Excluding the coating obtained in supporting electrolyte, all coatings show a small amount of Ce which is, except for the 13X+Ce sample, comparable to the limit of detection. Therefore, for all samples other than 13X+Ce, Ce concentration is questionable, especially if taking into account that EDS data were obtained with long integration time (180 seconds, 200x250 µm area of analysis).

Since we were not able to quantify the amount of Ce present in the PEO coating by EDS, we have made an attempt to make a qualitative presence of Ce by utilizing photoluminescence. The trivalent $Ce^{3+}$ ion has been widely studied as an efficient activator in $Al_2O_3$ as a host material. $Al_2O_3$ is an attractive host for rare-earth ions since it has many interesting properties such as high transparency from ultraviolet to near infrared region, excellent mechanical strength, high thermal conductivity, good chemical and photochemical properties, high melting point, etc. $Ce^{3+}$ doped $Al_2O_3$ PEO coatings exhibit an intense PL emission in ultraviolet/visible spectral range with PL maximum positioned around 345 nm under 285 nm excitation due to the transitions of $Ce^{3+}$ ions from 5d excited state to the 4f ground state [40-42].

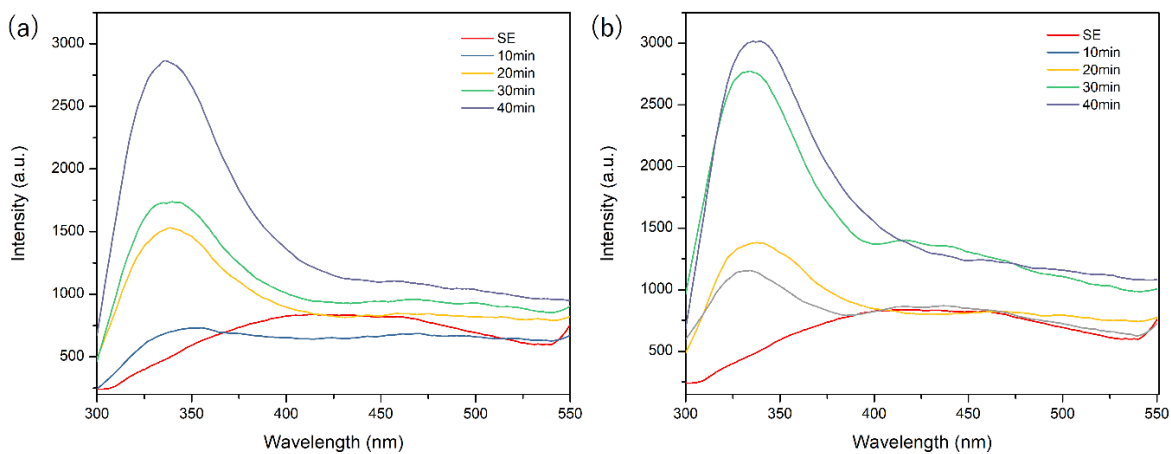

**Figure 5.** Emission PL spectra ($\lambda_{ex}$ = 285 nm) of obtained PEO coatings a) Cli+Ce; b)13X+Ce.

Emission PL spectra of PEO coatings obtained in electrolyte containing 1 g/L of 13x+Ce and Cli+Ce zeolites are presented in Fig. 5. PL spectra observed for coatings containing Ce-exchanged zeolites are composed of two spectral maxima. First sharp maximum is positioned at about 345 nm, while the other one is broad and it is centered at about 440 nm. By comparing emission PL spectrum of the coating obtained in supporting electrolyte without the addition of zeolite (SE in Fig. 5a and b), one can deduce that broad maximum is related to $Al_2O_3$ photoluminescence originating from optical transitions in PL centers which are defect centers related to oxygen vacancy defects in $Al_2O_3$ PEO coatings (F and $F^+$ centers) [43]. This finding is also in agreement with high resolution XRD patterns showing that coatings obtained for shorter PEO treatment times contain almost exclusively crystalline γ-$Al_2O_3$ phase (Fig. 2 b,c). Since PL intensity of $Ce^{3+}$ doped $Al_2O_3$ coatings depends on PEO processing time and concentration of Ce-species in electrolyte, i.e. it entirely depends on content of $Ce^{3+}$ ions incorporated into PEO coatings [40], we concluded that Ce is present in all coatings (with Ce concentration increasing with PEO time) but we were not able to quantitatively detect it because of its low content. Similar PL spectra were obtained for coatings obtained in electrolyte containing 1g/L of Ce-exchanged clinoptilolite, but Ce-related peak has a lower intensity due to lower concentration of Ce in the coatings, as a consequence of lower content of Ce in clinoptilolite.

*3.2 Photocatalytic activity of obtained coatings*

Photocatalytic activity of methyl orange photodegradation is presented in Fig. 6a for all obtained coatings. The best photocatalytic activity is observed for the coating obtained after 30 min of PEO processing in supporting electrolyte with addition of 1 g/L of 13X+Ce zeolite. In fact, one can observe that, for all coatings under investigation, photocatalytic activity increases with PEO treatment time up to 30 min, and then it slowly decreases. It is also clear that Ce-loading increases the photocatalytic activity for about 5 % for clinoptilolite and for about 10 % for 13X zeolite based samples, regardless of the fact that the concentration of Ce in 13X zeolite is about 10 times higher than in clinoptilolite. This suggests that clinoptilolite as a less expensive and natural zeolite may be used in photocatalysis, but it requires longer irradiation time compared to 13X zeolite (Fig 6b).

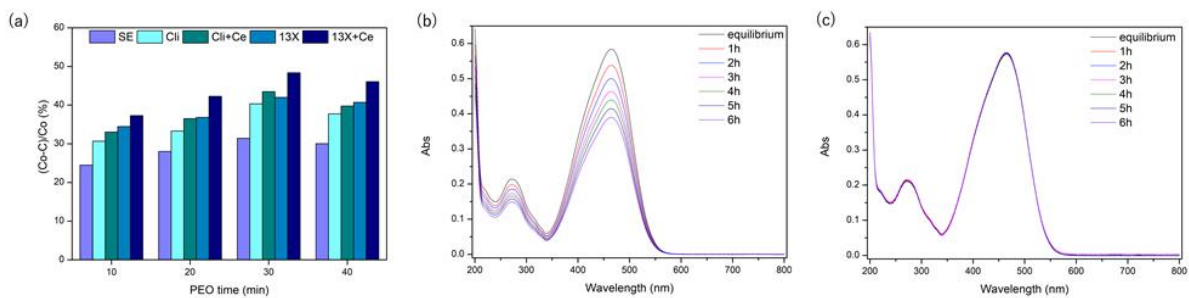

**Figure 6.** a) Photocatalytic degradation of MO on immobilized zeolite coatings; b) UV-Vis spectra of MO after various irradiation times in the presence of Cli+Ce sample; c) UV-Vis spectra of MO in the dark on zeolites immobilized on Al support.

In order to determine if there is an adsorption of methyl orange on photocatalysts, we conducted adsorption experiments, i.e. measurement of change in methyl orange concentration when samples are not illuminated (Fig. 6c). It's obvious that the adsorption is negligible and methyl orange photodegradation is the result of photocatalytic reactions in the presence of zeolites immobilized on aluminum support by plasma electrolytic oxidation.

To further probe optical properties of PEO coatings obtained by immobilizing zeolites on aluminum support, we performed a set of DRS measurements presented in Fig. 7a. Since it is not easy to observe absorption edge shift in DRS spectra we presented them as Tauc plots for coatings with immobilized 13X+Ce (Fig. 7b). Clearly, that bandgap of formed semiconducting oxide shrinks with PEO processing time, thus favoring the photocatalysis under longer irradiation wavelength, which is of particular interest [44].

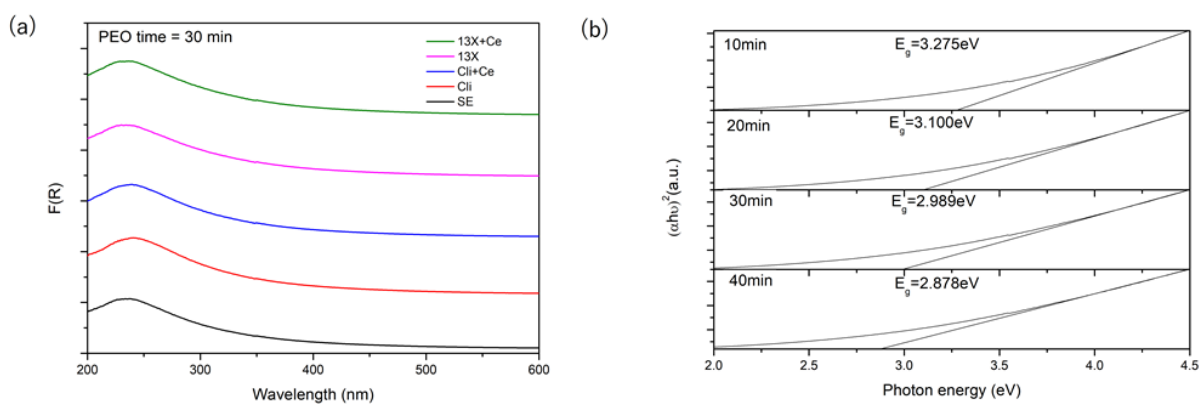

**Figure 7.** a) DRS spectra of coatings with immobilized zeolites processed for 40 min; b) Tauc plots of the coatings with 13X+Ce processed for various time.

Having in mind everything said above concerning immobilized zeolite coatings under investigation, it turns out that roughness, porosity, and optical properties influence their photocatalytic performance in methyl orange photodegradation. Unquestionably, the chemical composition of the coatings changes as well, which is indirectly shown by the increase of Ce-related PL emission maxima. All obtained coatings show an increase of photoactivity up to 30 min of PEO processing, while after this processing time photoactivity of the coatings decreases. This can be related to the fact that roughness of the coatings (Table 2), emitted PL intensity (Fig 5) (i.e., Ce concentration), and bandgap values (Fig 7b) do not change significantly after 30 min of PEO processing. Thus, at longer treatment times, surface area accessible for photocatalysis, number of photocatalytically active sites and energy required for the formation of electron-hole pairs in the semiconducting oxide do not change much anymore and thus do not affect the photocatalytic activity as much as at shorter PEO processing times. Porosity analysis shows that after 30 min of processing porosity increases which is related to the appearance of larger pores resulting from the appearance of smaller number of larger micro-discharges. Another thing that cannot be excluded from the discussion about photocatalytic properties is the decrease of high surface area γ-alumina phase [45] in the coatings and its possible covering by sillimanite for longer PEO processing times, as evidenced by XRD (Fig 2b). All these factors influence the photocatalytic activity and cause its decrease (or saturation) after 30 min of PEO processing.

On the other hand, the results obtained in this study are promising due to immobilization of Ce-exchanged and pure zeolites on aluminum support produces photocatalysts that are comparable to coatings that contain $TiO_2$, which is undoubtedly the most widely used photocatalyst used for photodegradation of organic pollutants (for example see Table 2 in Ref [46]).

*3.3 Degradation stability of obtained coatings*

In order to understand the role of Cli and 13X zeolites with and without $Ce^{3+}$ loading for the degradation behavior of the obtained coating, EIS measurements were performed. Fig. 8 shows the representative data for all the coatings obtained as a result of 20 minutes PEO treatment and respective bare material for comparison. The obtained results demonstrate certain scattering at low frequencies, which can be associated with active-passive transitions in the case of PEO coated substrates. The activation of a defect during the recording the impedance spectra causes drop of the low frequency impedance and can cause shift of the OCP and as a consequence a non-stationarity during recording of the data. Due to this fact, a detailed analysis of the EIS data was not performed. Nevertheless, the deep drops of the low frequency impedance are not typical for the Ce-containing systems evidencing the easier passivation of the formed defects. Comparing the evolution of impedance modulus curves for the zeolite-loaded PEO coatings with and without Ce, an important difference can be observed, namely a higher stability of the impedance modulus values at low frequencies for Ce-loaded systems. This effect might be attributed to an active corrosion inhibition by the cerium species.

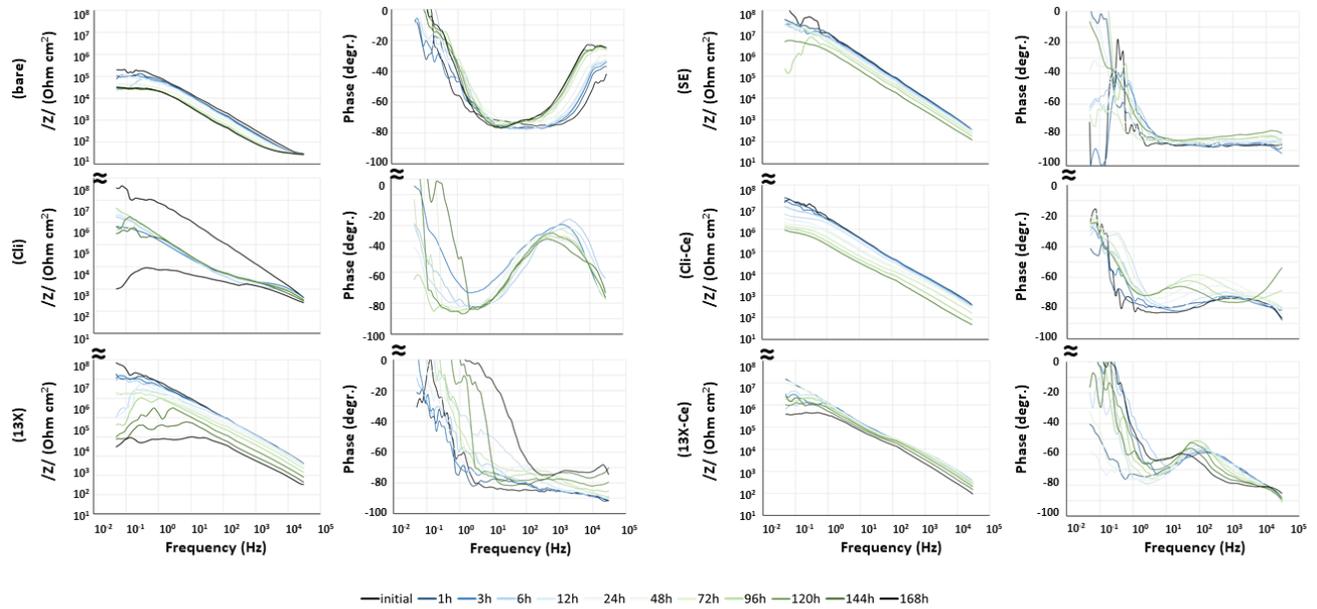

**Figure 8.** Representative EIS results for the PEO coatings obtained after 20 min PEO treatments. Bare material is shown for the comparison.

The most pronounced improvement can be observed for the 13X+Ce samples. This improvement is also confirmed by visual observation of the coatings after the EIS measurements (Table 4).

**Table 4.** Visual appearance of the surface after EIS measurements.

|        | SE  | Cli | Cli+Ce | 13X | 13X+Ce |
|--------|-----|-----|--------|-----|--------|
| 10 min | - - | -   | +      | -   | ++     |
| 20 min | -   | -   | -      | C   | ++     |
| 30 min | C   | - - | +      | +   | ++     |
| 40 min | -   | -   | +      | ++  | +      |

"- -": large area of coating detachment; "-": small localized detachment of coating; C – visible corrosion products; "+": healed corrosion defects (passive-active-passive transition); "+ +": no visible corrosion signs

Different kind of defects were observed on the surface after the corrosion test. In general, they can be divided in three groups (Fig 9):

1) Initial pitting type of defect in the coating (Fig. 9a and 9c). Comparing with EIS measurements, one can assume, that those defects are not crucial, and do not lead to a significant drop of impedance. In other words, those pittings, which are formed at the beginning of exposure to sodium chloride media, are not stable and undergo active-passive transitions [ref], and therefore do not lead to a strong drop of low frequency impedance.

2) Detachment of coating areas (Fig. 9b). Larger or smaller zones of PEO coatings are cracked and (partially) flaked-off. These damages show significant drops in impedance modulus as observed during EIS measurements and visual inspection of samples.

3) Corrosion products formation (Insert in Fig. 9c), which are also not crucial for the corrosion behavior obviously due to sealing of coating porosity and high substrate resistance by itself.

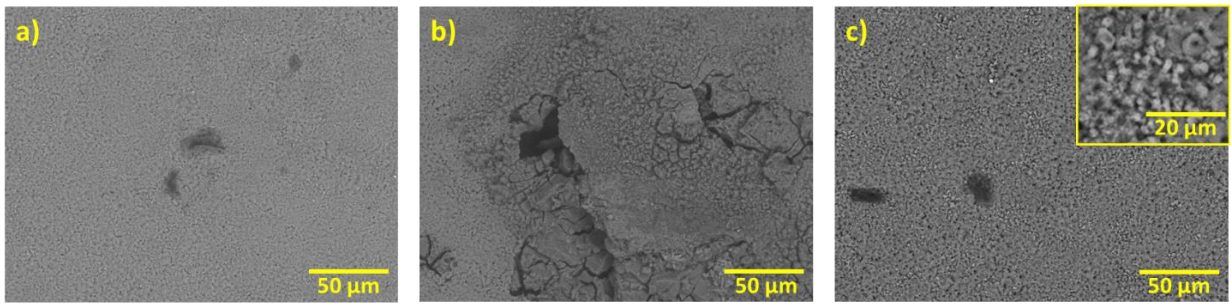

**Figure 9.** Classification of defects observed on the coating surface after EIS measurements.

In spite of the defect formation and presence of corrosion products, one can see the overall improvement of coatings after loading zeolites with $Ce^{3+}$. This can be explained via the formation of insoluble Ce(III) and Ce(IV) hydroxide precipitates blocking the active defects especially at the cathodic areas [47,48].

Overall, the effect of Ce is not too strongly pronounced due to its low concentration in the coating, as demonstrated by EDS measurements and high corrosion resistance of substrate itself and its ability to passivate, but still visible during corrosion testing especially with better long term stability in the case of 13X+Ce.

## 4. Conclusions

Oxide coatings with immobilized pure and Ce-exchanged zeolites are deposited on aluminum support from silicate-based electrolyte using PEO processing. Obtained coatings are characterized for their morphology, chemical and phase composition, photocatalytic activity and anti-corrosion properties. The following conclusions are drawn:

1) All coatings are partially crystallized containing γ-alumina and aluminosilicate phase sillimanite: γ-alumina crystalline phase is dominant for lower PEO processing times and it probably gets suppressed by sillimanite in the later stages of processing.

2) Surface morphology of obtained coatings is independent of the used zeolite, but it strongly depends on PEO processing time: roughness of all coatings increases with PEO time, as well as the number of pores and their size, while total porosity reaches its minimum after 30 min of PEO treatment.

3) All coatings contain elements originating from the substrate and from the electrolyte. Cerium concentration is very low in all coatings, as revealed by EDS. Higher EDS integration times and larger areas under investigation show that the ratio of Ce in Cli+Ce and 13X+Ce is comparable to the ratio obtained by the chemical analysis of zeolite powders. We have used PL as a tool to qualitatively probe Ce content in the coatings and showed that it increases with prolonged PEO processing time.

4) The highest photocatalytic activity is observed for coating with immobilized Ce-exchanged 13X zeolite processed for 30 min. Coatings containing zeolites with Ce show higher photoactivity then those with immobilized pure zeolites. All four series of coatings show an increase of photoactivity up to 30 min of processing and its drop later on. This behavior is related to Ce content, coating's morphology, estimated semiconducting oxides' bandgap value, and topping of γ-alumina by sillimanite.

5) The effect of Ce is not too strongly pronounced on anti-corrosion properties due to its low concentration in the coatings but it is still visible. Corrosion observations suggest that coatings containing Ce-exchanged 13X zeolite have the highest corrosion resistance which lapses for higher PEO processing time.

**Conflicts of interest**

There are no conflicts of interest to declare.

**Acknowledgements**

The authors acknowledge the work of Prof. dr Dr. Vladislav Rac from the Faculty of Agriculture of the University of Belgrade in obtaining DRS spectra. This work is supported by the Ministry of Education, Science, and Technological Development of the Republic of Serbia (SS and LjDV acknowledge the financial support by MESTD Contract number: 451-03-9/2021-14/200146) and by the European Union Horizon 2020 research and innovation program under the Marie Sklodowska-Curie grant agreement No. 823942 (FUNCOAT).

**CRediT**

Kristina Mojsilović (Investigation, Data curation, Visualization), Nikola Božović (Investigation, Data curation), Srna Stojanović (Investigation), Ljiljana Damjanović-Vasilić (Conceptualization, Methodology, Writing), Maria Serdechnova (Investigation, Writing), Carsten Blawert (Writing, Supervision), Mikhail L. Zheludkevich (Supervision), Stevan Stojadinović (Methodology), Rastko Vasilić (Conceptualization, Supervision, Writing – Reviewing and Editing).